\documentstyle[12pt,epsfig]{article}

\newcommand{\abstracts}[1]{{
\centering{\begin{minipage}{12.2truecm}
\normalsize\baselineskip=15pt
\centerline{\footnotesize ABSTRACT}\vspace*{0.3cm}
\parindent=20pt #1
\end{minipage}}\par}}

\newcommand{\dd}{\mbox{\rm d}}

\newcommand{\cD}{{\cal D}}
\newcommand{\beqn}{\begin{eqnarray}}
\newcommand{\eeqn}{\end{eqnarray}}
\newcommand{\eq}[1]{(\ref{#1})}

\begin{document}
\pagestyle{myheadings}
\thispagestyle{empty}
\markright{JETP Lett., Vol. 66, No. 9, 10 Nov 1997, pp.605-608.}
\setcounter{footnote}{0}
\renewcommand{\thefootnote}{\fnsymbol{footnote}}

\begin{center}

{\baselineskip=16pt
{\Large \bf On Nambu Monopole Dynamics in the ${\mathbf{SU(2)}}$
Lattice Higgs Model}\\

\vspace{1cm}

{\large M.~N.~Chernodub\footnote{e-mail: chernodub@vxitep.itep.ru}}\\

\vspace{.5cm}
{\it Institute of Theoretical and Experimental Physics,\\
B.~Cheremushkinskaya 25, Moscow, 117259, Russia}}
\end{center}

\vspace{1cm}

\abstracts{
It is shown that the $SU(2)$ Higgs model on a lattice is equivalent to
the Georgi--Glashow model in the limit of a small coupling constant
between the Higgs and gauge fields. It can therefore be concluded
that the transition between the confinement and symmetric phases in the
$3+1$ dimensional $SU(2)$ Higgs model at finite temperature is
accompanied by condensation of Nambu monopoles.}

\vspace{1cm}

According to one of the most popular modern hypotheses, the baryon
asymmetry of the universe may arise in the process of an electroweak
phase transition (see, for example, the review by Rubakov and
Shaposhnikov~\cite{RuSh96}).

Due to the smallness of the Weinberg angle $\theta_W$ and the
insignificance of the fermion effects, this transition is largely
determined by the properties of the $SU(2)$ Higgs model. The present
letter examines the behavior of the magnetic fluctuations which can
play an important role in a temperature phase transition in the
SU(2) Higgs model.

Let us consider the SU(2)
lattice Higgs model with scalar field $\Phi_x$ in the fundamental
representation, the action is:
\beqn
S[U,\Phi] = - \frac{\beta}{2} \sum_P {\rm Tr} U_P - \frac{\kappa}{2}
\sum_x \sum_\mu \Bigl(\Phi^+_x U_{x,\mu} \Phi_{x+\hat\mu} +
c.c.\Bigr) + V(|\Phi|)\,,
\label{S}
\eeqn
Here $U_P$ represents the ordered product of the edge elements of the
gauge field $U_{x,\mu}$ over the boundaries faces of the plaquette
$P$, and $V(\Phi)$ is the potential on the field $\Phi$, and
${|\Phi|}^2 = \Phi^+ \Phi$.

Due to the triviality of the homotopy group $\pi_2 (SU(2))$, there
are no topologically stable monopole defects in this theory.
However, ``embedded''~\cite{VaBa92} monopoles, the so-called ``Nambu
monopoles''~\cite{Na77}, which are not topologically stable defects,
do exist in the theory.  These objects are described by the composite
field
\beqn
\chi^a_x = \Phi^+_x \sigma^a \Phi_x\,,
\label{chi}
\eeqn
($\sigma^a$ are the Pauli matrices) which behaves under gauge
transformations as a scalar field in the adjoint representation. A
Nambu monopole is a configuration of fields $U$ and $\Phi$ such that
the field $U$ and the composite field $\chi$, expressed in terms of
the fundamental field $\Phi$, according to~\eq{chi}, coincide
with the
configuration of the 't~Hooft--Polyakov monopole~\cite{tHPo74} in the
Georgi--Glashow model~\cite{GeGl72} with the field $\chi$ in the
adjoint representation and with the gauge field $U$. Since Nambu
monopoles are described solely by the gauge field $U$ and the
composite field $\chi$, the dynamics of these monopoles is determined
completely by the effective action $S_{eff}$:
\beqn
e^{- S_{eff}[U,\chi]} = \int \cD \Phi \, e^{- S[U,\Phi]}
\prod_a \prod_x \delta(\chi^a_x - \Phi^+_x \sigma^a \Phi_x)\,.
\label{eff0}
\eeqn
To calculate the action $S_{eff}$ it is convenient to study the
following parametrization of the field $\Phi$:
\beqn
\Phi = e^{i \varphi} \, \Psi\,,\qquad
\Psi = \rho \, {\cos \alpha \, e^{i \theta}\choose
\sin\alpha}\,, \nonumber
\eeqn
where $\varphi,\theta \in [-\pi,\pi)$, $\alpha\in [0,\pi \slash 2]$
and $\rho \in [0,+\infty)$.
The fields $\rho$,
$\alpha$ and $\theta$ can be
expressed in terms of the field $\chi$ with the aid of Eq.\eq{chi}!:
\beqn
\theta = \arctan \frac{\chi^2}{\chi^1}\,,
\qquad \alpha & = & \frac{1}{2} \arctan
\frac{\sqrt{{(\chi^1)}^2 + {(\chi^2)}^2}}{|\chi^3|}\,,
\qquad \rho = \sqrt{|\chi|}\,,\nonumber
\eeqn
whence
\beqn
\Psi = \frac{1}{\sqrt{2 \, (|\chi| - |\chi^3|)}}
\cdot {\chi^1 + i \chi^2
\choose |\chi| - |\chi^3|}\,.
\label{Psi}
\eeqn
Using the relation for the modulus of the field $\Phi$, ${|\Phi|}^2 =
|\chi| = \sqrt{(\sum\limits^3_{a=1} {(\chi^a)}^2 )}$,
and the measure
\beqn
\int\limits^{+\infty}_{-\infty} \cD \Phi \, \cdots =
\int\limits^\pi_{-\pi} \cD \varphi \cdot
\int\limits^{+\infty}_{-\infty}
\prod_x  \frac{1}{|\chi_x|} \dd \chi^1_x
\dd \chi^2_x \dd \chi^3_x\, \cdots\,, \nonumber
\eeqn
we get for the effective action \eq{eff0}:
\beqn
S_{eff}[U,\chi] = - \frac{\beta}{2} \sum_P
{\rm Tr} U_P + S_h [U,\chi] + {\tilde V}(|\chi|)\,,
\label{Seff1}
\eeqn
where the new potential on the field $\chi$ is determined by
the expression
\beqn
{\tilde V}(|\chi|) = V(\sqrt{|\chi|})
+ \sum_x \ln |\chi_x|\,,
\eeqn
and the interaction of the fields $U$ and $\chi$ is
\beqn
e^{- S_h [U,\chi]} = \int\limits^\pi_{-\pi} \cD \varphi \exp\Bigl\{
\kappa \sum_x \sum_\mu R_{x,\mu} \cos(\varphi_{x+\hat\mu}
- \varphi_x + A_{x,\mu}) \Bigr\}\,.
\label{Sh}
\eeqn
In this formula we introduced the notation
\beqn
\Psi^+_x U_{x,\mu} \Psi_{x+\hat\mu} = R_{x,\mu} e^{i A_{x,\mu}}\,.
\label{R}
\eeqn
The derivation of the effective action~\eq{Seff1} is correct in any
dimension of space--time.

For simplicity, we shall examine the case of an infinitely deep
potential $V(|\Phi|)$ with the minimum ${|\Phi|}^2 = |\chi| = 1$.  In
this case, the lengths of the Higgs field $\Phi$ and of the composite
field $\chi$ are frozen. The integral~\eq{Sh} is most easily
calculated in the limit~$\kappa \ll 1$.  In leading order we obtain
(to within an additive constant):
\beqn
S_h = - \frac{\kappa^2}{2} \sum_x \sum_\mu R^2_{x,\mu}
+ O(\kappa^4) = - \frac{\kappa^2}{8} \sum_x \sum_\mu
{\rm Tr} (U_{x,\mu} \, \chi_x \, U^+_{x,\mu} \, \chi_{x+\hat\mu})
+ O(\kappa^4)\,, \nonumber
\eeqn
where we employed Eqs.\eq{Psi} and \eq{R} and introduced the notation
$\chi = \chi^a \sigma^a$. Thus in the limit $\kappa \ll 1$  the
effective action \eq{Seff1} with the length ${|\Phi|}^2 = 1$ of the
Higgs field frozen is identical in leading order to the
Georgi--Glashow action:
\beqn
S_{eff}[U,\chi] = - \frac{\beta}{2} \sum_P
{\rm Tr} U_P - \frac{\gamma}{2} \sum_x \sum_\mu {\rm Tr} (U_{x,\mu}
\, \chi_x \, U^+_{x,\mu} \, \chi_{x+\hat\mu}) + O(\kappa^4)\,,
\label{GG}
\eeqn
where
\beqn
\gamma = \frac{\kappa^2}{4}\,. \label{gamma}
\eeqn

It is interesting to compare the phase diagrams of the
$3+1$--dimensional $SU(2)$ Higgs model \eq{S} and the Georgi--Glashow
model (\ref{GG},\ref{gamma}) at nonzero temperature with radially frozen
Higgs fields. Figure~1
displays schematically the phase diagram obtained in
Ref.~\cite{EvJeKa86} for the $SU(2)$ Higgs model.  Figure~2 shows the
phase diagram obtained in Ref.~\cite{KaSeSt83} for the Georgi--Glashow model.
For small values of the constant $\beta$ both theories are in the
confinement phase (color confinement). As $\beta$ increases, a
phase transition from the confinement phase to the symmetric phase
occurs in both theories at small $\kappa$.  The line of phase transitions
$A^\prime$-$B^\prime$ in the Georgi--Glashow model should correspond
to the line of phase transitions $A$-$B$ in the $SU(2)$ Higgs model
according to Eq.~\eq{gamma}:  $\gamma_c (\beta) = \kappa^2_c (\beta)
\slash 4 + O(\kappa^4_c)$.

Figure~2 shows schematically the phase transition predicted with the
aid of Eq.~\eq{gamma} (dashed line $A^\prime$-$C^\prime$).
Unfortunately, it is impossible to determine the correctness of this
prediction quantitatively on the basis of the results of
Refs.~~\cite{EvJeKa86} and~\cite{KaSeSt83}, since in those papers the
phase diagrams were studied at different temperatures.

It is known~\cite{Po77} that in the Georgi--Glashow model confinement
is due to the dynamics of the 't Hooft--Polyakov monopoles: in the
confinement phase the monopoles are condensed, while in the
deconfinement phase there exists a dilute gas of
monopole--anti-monopole pairs. Therefore, at least for low values of
the coupling constant $\kappa$, the phase transition from the
symmetric phase to the confinement phase in the $SU(2)$ Higgs
model~\eq{S} is accompanied by condensation of Nambu monopoles, since
the Nambu monopoles in the $SU(2)$ Higgs model~\eq{S} correspond to the
't~Hooft--Polyakov monopoles in the Georgi--Glashow model~\eq{GG}.

It is natural to suppose that condensation of Nambu monopoles also occurs
for larger values of the parameter $\kappa$ for the phase
transitions from the confinement phase to the symmetric phase and from the
confinement phase to the Higgs phase. The latter conjecture finds support
in the fact that in the Higgs phase there exists an embedded string, the
so-called $Z$-vortex string~\cite{Na77,Ma83}, with nonzero string tension.
Stretched between Nambu monopoles, such a string destroys the condensate.
The results of investigations of this question will be published later in
a separate paper. 

I am grateful to E.--M.~Ilgenfritz and M.I.~Polikarpov for helpful
remarks. This work was supported in part by the Russian Fund for
Fundamental Research under Grant No.~96-02-17230a and under the
Grants INTAS-94-0840 and INTAS-RFBR-95-0681. I also thank the Japan
Society for the Promotion of Science (JSPS) for financial assistance,
provided as part of the program for the support of scientists of the
former Soviet Union.

\section*{Note Added}

The dynamical role of these embedded topological defects has been further
studied in Refs.~\cite{10,11,12} in a $3D$ $SU(2)$ model which is an
effective dimensionally reduced version of the Electroweak
theory~\cite{13,14,15}. The $Z$--vortices exhibit a percolation transition
which accompanies the first order thermal transition at small Higgs
mass~\cite{10}. The percolation transition exists also in the cross-over
regime at high Higgs masses~\cite{11}.  Ref.~\cite{12} gives a short
review of these and other related results.

\newpage

\section*{Figures}

\begin{figure}[!htb]
\centerline{\epsfig{file=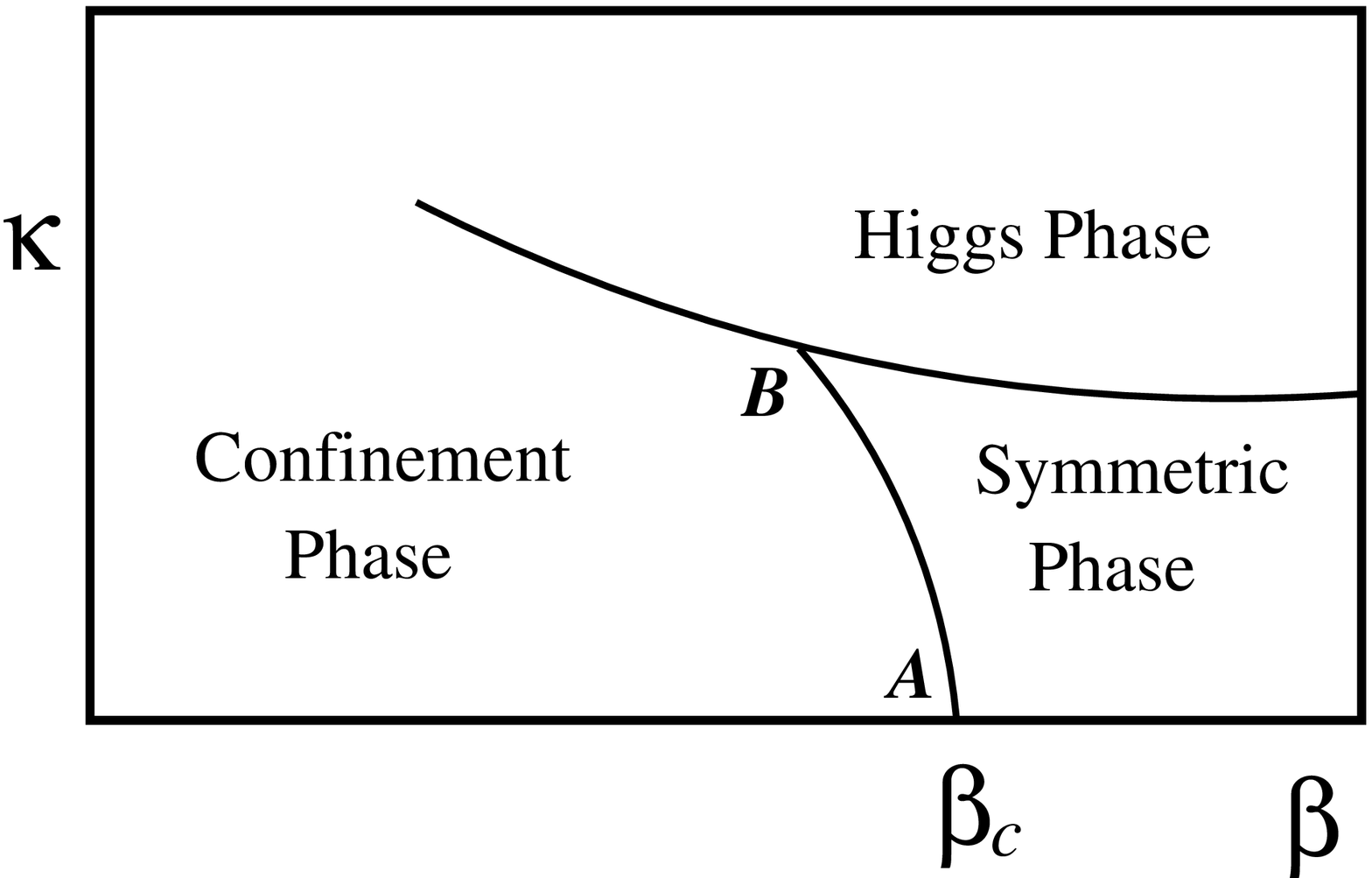,height=6.0cm}}
\caption{Phase diagram of the $SU(2)$ Higgs model at nonzero temperature, 
the Higgs field radius is fixed.}
\end{figure}

\begin{figure}[!htb]
\centerline{\epsfig{file=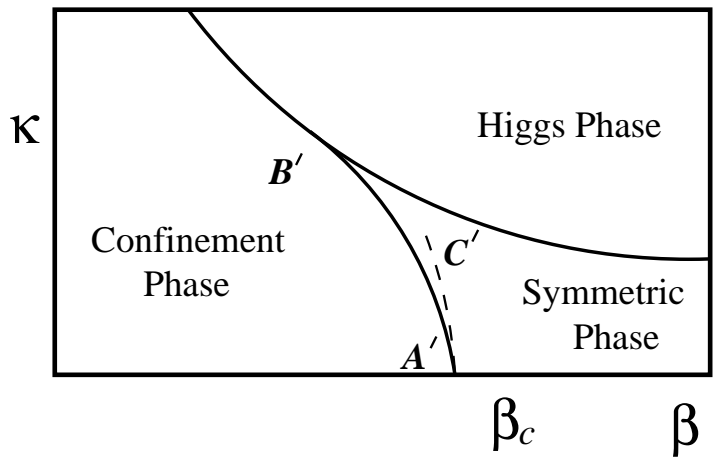,height=6.0cm}}
\caption{The same as in Figure~1 but for the Georgi--Glashow model.
}
\end{figure}

\end{document}